\documentclass[10pt,aps,pra,twocolumn]{revtex4-2}
\usepackage{amsmath}
\usepackage{amssymb}
\usepackage{graphicx}
\usepackage{xcolor}
\usepackage{hyperref}
\hypersetup{
    colorlinks=true,
    linkcolor=blue,
    citecolor=blue,
    filecolor=blue,      
    urlcolor=blue,
}

\DeclareMathOperator{\sinc}{sinc}

\begin{document}

\title{Correlated self-heterodyne method for ultra-low-noise laser linewidth measurements}

\author{Zhiquan Yuan$^{1,\dagger}$, Heming Wang$^{1,\dagger}$, Peng Liu$^{1,2,\dagger}$, Bohan Li$^{1}$, Boqiang Shen$^{1}$, Maodong Gao$^{1}$, Lin Chang$^{3}$, Warren Jin$^{3}$, Avi Feshali$^{4}$, Mario Paniccia$^{4}$, John Bowers$^{3}$ and Kerry Vahala$^{1,*}$\\
$^{1}$T. J. Watson Laboratory of Applied Physics, California Institute of Technology, Pasadena, CA, USA\\
$^{2}$School of Electronics Engineering and Computer Science, Peking University, Beijing 100871, China\\
$^{3}$ECE Department, University of California Santa Barbara, Santa Barbara, CA, USA\\
$^{4}$Anello Photonics, Santa Clara, CA, USA\\
$^{\dagger}$These authors contributed equally.\\
$^{*}$vahala@caltech.edu}

\begin{abstract}
Narrow-linewidth lasers are important to many applications spanning precision metrology to sensing systems. Characterization of these lasers requires precise measurements of their frequency noise spectra. Here we demonstrate a correlated self-heterodyne (COSH) method capable of measuring frequency noise as low as $0.01$ Hz$^2$/Hz at 1 MHz offset frequency. The measurement setup is characterized by both commercial and lab-built lasers, and features low optical power requirements, fast acquisition time and high intensity noise rejection.
\end{abstract}

\maketitle

\section{Introduction}
Ultra-low-noise lasers are indispensable ingredients for a wide range of applications, including optical gyroscopes \cite{lai2020earth}, optical atomic clocks \cite{newman2019architecture}, and light detection and ranging (LiDAR) systems \cite{suh2018soliton}. Accurate measurement of ultra-low-noise frequency spectra is an essential prerequisite for optimizing their performance and advancing their applications. Hence, high demands are placed on measurement systems to characterize these lasers with low frequency noise floors and high intensity-fluctuation isolation. 

Several methods have been used for laser linewidth characterization. Incoherent homodyne detection incorporates a fiber delay line exceeding the coherence length of the laser under test \cite{schumaker1984noise}. While the method measures relatively high frequency noise levels accurately, it becomes inappropriate as the laser linewidth reaches Hz levels, where the coherence length is on the order of $10^5$ km. Phase discriminators with sub-coherent-length delay have been proposed in such cases, either by locking to a quadrature point \cite{li2012characterization,lee2012chemically} or using self-heterodyne detection to shift the signal to the radiofrequency (RF) domain so as to avoid low-frequency technical noise \cite{van1992excess,ludvigsen1998laser}. However, optical-to-electrical (OE) conversion at the photodetector (PD) introduces additional technical PD noise, and relative intensity noise (RIN) of the laser may also be coupled to the output signal through the OE conversion. These factors prevent the detection methods from achieving the sufficiently-low noise floor required for milli-Hertz-linewidth laser characterization. At the same time, cross-correlation has been applied to frequency noise characterization in the RF domain as a mature method for measuring ultra-low-noise microwave signals \cite{walls1992cross}. This technique compares the signal against two references and correlates them to suppress the independent noise from the references. Similar techniques have been introduced in the optical domain to characterize sub-Hertz linewidth lasers with optical references \cite{xie2017phase}.

In this paper, we demonstrate a reference-free self-heterodyne cross-correlator for ultra-low-noise laser linewidth measurements. By employing two PDs for OE conversion in the conventional self-heterodyne method, the cross-correlation can eliminate the need for references while extracting laser noise. In addition, balanced photodetectors (BPDs) are used to minimize the RIN coupling ratio. The cross-correlator has a noise floor lower than 0.01 $\text{Hz}^2$/Hz and 40.4 $\text{dBrad}^2$ RIN suppression at 1 MHz offset frequency. Various commercial and lab-built lasers are used to benchmark the measurements, including an external-cavity laser (ECL) and a distributed-feedback (DFB) laser with and without self injection locking. Factors that may impact cross-correlator performance, including environmental noise coupling and delay length selection, are also discussed. The measurement can be readily generalized to other wavelengths \cite{tran2021extending} and may advance the development of next-generation laser sources through rapid measurement of noise.

\section{Measurement setup}

\begin{figure*}
\includegraphics[width=132mm]{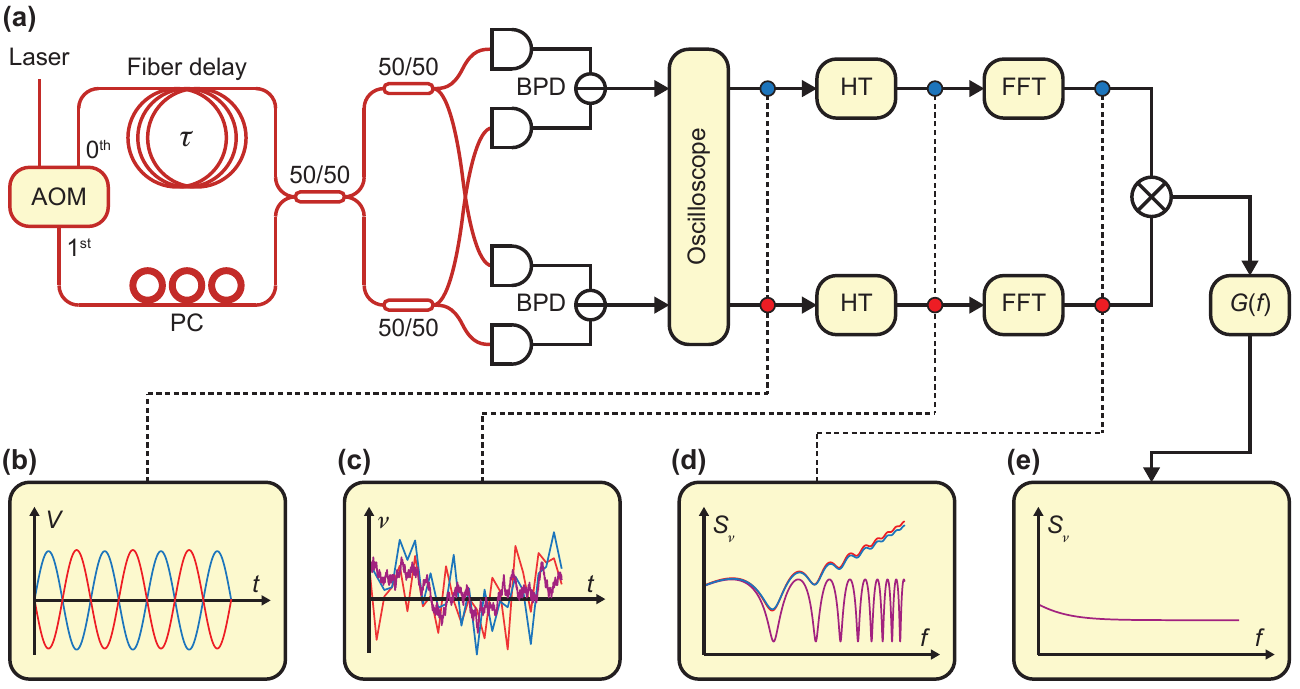}
\caption{Schematic of the correlated self-heterodyne (COSH) measurement setup.
(a) Illustration of the setup. AOM, acousto-optic modulator; PC, polarization controller; BPD, balanced photodetector; HT, Hilbert transform, followed by a time-domain difference operation; FFT, fast Fourier transform.
(b) Self-heterodyne waveforms recorded by a high-speed oscilloscope.
(c) Frequency fluctuations extracted with Hilbert transform and time-domain difference. The red and blue lines refer to the frequency fluctuations from two BPD outputs marked by color in accordance with inset (a) (similar hereinafter). Laser frequency noise (common mode) indicated by the purple line and BPD-induced noise (differential mode) are contained in the extracted results.
(d) Power spectral density (PSD) of BPD output frequency given by FFT. The red and blue lines refer to the total noise PSD while the purple line indicates the laser frequency noise PSD (common mode). The deviation between them at high frequency is due to the BPD noise.
(e) Single-sideband (SSB) laser noise after cross-correlation and $G(f)$ in which the BPD noise has been eliminated.}
\label{FIG1}
\end{figure*}

The COSH setup is shown in Fig. \ref{FIG1}(a). A laser under test is split by a three-port acousto-optic modulator (AOM) into frequency-downshifted ($1^{\text{st}}$ order output) and unshifted ($0^{\text{th}}$ order output) portions. The former is polarization-controlled and then recombined with the latter delayed by a 1-km-long fiber, which forms a modified Mach–Zehnder interferometer (MZI) with a free spectral range (FSR) of 214.06 kHz. The three-port AOM acts as a variable splitter that also uses the unshifted laser power compared to a two-port modulator. On the output side, instead of one PD, both outputs are divided and received by two identical balanced photodetectors (BPDs). Using BPDs helps suppress RIN and using two BPDs allows cross-correlation between the electrical outputs and suppresses independent BPD noise. The whole optical section of the system is isolated from the external environment with an acoustic shield. The BPD outputs are recorded using a high-speed oscilloscope. The AOM is driven with a 55 MHz radio-frequency carrier which determines the center frequency of the recorded waveforms [Fig. \ref{FIG1}(b)], and the sampling rate of the oscilloscope is set to 250 MHz to prevent aliasing. AC coupling at the oscilloscope is used to block low-frequency components that are spectrally far away from the carrier. The time delay between the two channels is estimated to be less than 0.5 ns and will not be considered in the following analyses. 2 seconds of waveforms ($500 \times 10^6$ points for each channel, $1 \times 10^9$ points in total) are collected and transferred to a computer for data processing. While the record length is limited by the memory of our oscilloscope, it is sufficient to meet the noise floor requirements for the current measurements (see Section 4.3).

In the data processing part, phase fluctuations for each channel are extracted using Hilbert transforms and then converted to frequency fluctuation through time-domain difference [Fig. \ref{FIG1}(c)]. The Hilbert transform causes distortions at the endpoints of the waveforms, thus the first and last 40 ms ($10 \times 10^6$ points) for each channel are discarded after the Hilbert transform. The remaining points are divided into non-overlapping segments (rectangular windowing, similar to Bartlett's method for estimating power spectra \cite{bartlett1948smoothing}), each with a $\tau_R$ time length [corresponding to a resolution bandwidth (RBW) of $1/\tau_R$] and fast-Fourier-transformed to obtain the frequency spectrum at a given offset frequency. For lower frequency offsets, the RBWs are made smaller with the segment lengths adjusted accordingly. The power spectrum density (PSD) of BPD output frequencies suffer from BPD noise [Fig. \ref{FIG1}(d)]. To suppress this noise, the cross-correlation is calculated as the product between the Fourier coefficients of the two BPD output frequency spectra, averaged over all available segments. Owing to the independent nature of the two BPD noise sources, BPD noise can be suppressed compared to individual PSDs of BPD outputs. Finally, the cross-correlation spectrum is multiplied by a processing gain $G(f)$ to compensate the filtering effect from the MZI and recover the single-sideband (SSB) laser frequency noise result [Fig. \ref{FIG1}(e)].
 
% Theory
\section{Modeling of the measurement process}

In this section, we model the properties of the output signal from the COSH measurements. In the following, all spectral densities refer to two-sided spectral densities unless indicated otherwise.

\subsection{Self-heterodyne noise detection}
To model the self-heterodyne method for noise measurement, we begin with an idealized derivation, where a frequency-shifted signal beats against a delayed signal and the resulting frequency noise is calculated. From the Wiener-Khinchin theorem, the frequency noise of the original signal can be understood as the Fourier transform of the phase time-derivative correlation function:
\begin{align}
S_\nu(f) &=\int_{-\infty}^{\infty} \left<\frac{\dot{\phi}(0)}{2\pi}\frac{\dot{\phi}(t)}{2\pi}\right> \mathrm{e}^{2\pi ift}\mathrm{d}t \nonumber\\
&=\frac{1}{4\pi^2}\int_{-\infty}^{\infty} \left<\dot{\phi}(0)\dot{\phi}(t)\right> \mathrm{e}^{2\pi ift}\mathrm{d}t
\end{align}
Here $S_\nu(f)$ is the frequency noise PSD of the input signal, $f$ is the offset frequency, $\phi$ is the optical noisy phase signal, $t$ denotes time, dot indicates derivative with respect to $t$ and $\left<z\right>$ denotes the ensemble average of $z$.

The self-heterodyne beating outputs a signal with phase $-2\pi f_\mathrm{c} t +\phi(t)-\phi(t-\tau)$, where $\tau$ is the delay time and $f_\mathrm{c}$ is the carrier frequency (determined by the AOM). We are thus interested in the frequency noise of $\nu(\tau)=-f_\mathrm{c}+[\dot{\phi}(t)-\dot{\phi}(t-\tau)]/(2\pi)$. By the time-shifting property of the Fourier transform we get
\begin{align}
S_{\nu(\tau)}(f) &=[2-\exp(2\pi if\tau)-\exp(-2\pi if\tau)]S_\nu(f) \nonumber\\
&=4\sin^2(\pi f\tau)S_\nu(f)
\end{align}
thus the frequency noise reads
\begin{equation}
S_{\nu(\tau)}(f)=4\sin^2(\pi f\tau)S_\nu(f)
\label{Eq_Snutau_simple}
\end{equation}
The transfer function in the above equations has zeros at integer multiples of MZI FSR ($f=1/\tau$, $2/\tau$, $3/\tau$ and so on). At these offset frequencies, destructive interference eliminates the phase difference at the MZI output. To compensate, the processing gain $G(f)$ should be chosen as
\begin{equation}
G(f)=\frac{1}{4\sin^2(\pi f\tau)}
\label{Eq_Gf_simple}
\end{equation}
We note that $G(f)$ diverges at integer multiples of the MZI FSR. This nonphysical divergence will be removed after the finite detection resolution bandwidth is properly considered (see discussion in section 3.3).

It is sometimes convenient to directly measure the phase noise corresponding to $S_{\nu(\tau)}(f)$, denoted as $S_{\varphi,\nu(\tau)}(f)\equiv S_{\nu(\tau)}(f)/f^2$ (e.g. by sending the RF signal directly to a phase noise analyzer). In this case,
\begin{equation}
S_{\varphi,\nu(\tau)}(f)=4\frac{\sin^2(\pi f\tau)}{f^2}S_\nu(f)=(2\pi\tau)^2 \sinc^2(f\tau)S_\nu(f)
\end{equation}
where $\sinc z\equiv\sin(\pi z)/(\pi z)$ is the normalized sinc function.

\subsection{Self-heterodyne noise detection using an AOM}
Here we present a more rigorous derivation based on the setup described previously. Various non-ideal effects can be incorporated and compared against the experiments.

We assume a laser input signal of the form
\begin{equation}
A(t)=\exp(-2\pi if_0 t)[1+\delta a(t)]\exp[-i\delta \phi(t)]
\end{equation}
where $f_0$ is the optical frequency. The $\delta a$ and $\delta \phi$ are relative amplitude fluctuation and phase fluctuation, respectively, and are assumed to be small within the time scale of $1/f_\mathrm{c}$, where $\delta \phi = \phi -2\pi f_0 t$. This indicates that the laser would have low noise, which is the intended regime for the cross-correlator. Discussions on using the setup to measure a high-noise laser can be found below in Section 4.2.

The frequency-shifted signal becomes
\begin{widetext}
\begin{equation}
A_1(t)=\frac{1}{\sqrt{2}}\exp(-2\pi if_0 t)\exp(2\pi i f_\mathrm{c} t)[1+\delta a(t)]\exp[-i\delta \phi(t)]
\end{equation}
and the delayed signal becomes
\begin{equation}
A_2(t)=\frac{1}{\sqrt{2}}\exp(-2\pi if_0 t)\exp(2\pi if_0 \tau)[1+\delta a(t-\tau)]\exp[-i\delta \phi(t-\tau)]
\end{equation}
where we have assumed that the AOM splits the light equally into two ports (which can be realized by adjusting the RF power input for the AOM).

The signals from the two arms are mixed at another coupler and form the MZI outputs. We do not assume \textit{a priori} that the coupler is perfectly balanced and write the two output amplitudes as
\begin{equation}
A_+=q_1A_1+iq_2A_2,\ \ A_-=q_1^*A_2+iq_2^*A_1
\end{equation}
where $q_1$ and $q_2$ are complex transmission coefficients. We further assume that the two couplers just before the BPDs are matched, such that the relative power between the two arms remains the same at the two BPDs. In this case, powers at individual PDs can be found from $P_\pm\equiv|A_\pm|^2$, and the RF power of each BPD output reads, up to a proportional constant,
\begin{equation}
\Delta P \equiv P_+-P_- = 2\mathrm{Re}[iq_1^*q_2\exp(2\pi i f_0 \tau)\exp(-2\pi i f_\mathrm{c} t)(1+\Sigma a)\exp(i\Delta \phi)] +\Delta|q|^2\Delta a
\end{equation}
where we have introduced some shorthand notations: $\Sigma a\equiv\delta a (t)+\delta a (t-\tau)$, $\Delta\phi \equiv \delta\phi(t)-\delta\phi(t-\tau)$, $\Delta a\equiv\delta a (t)-\delta a (t-\tau)$, and $\Delta |q|^2\equiv |q_1|^2-|q_2|^2$.

Next, the Hilbert transform is performed on $\Delta P$ to recover the analytic signal and extract the instantaneous phase. The terms within the brackets consist of the main part of $\Delta P$ and is itself an analytic signal oscillating at $f_\mathrm{c}$. The $\Delta a$ term may also influence the phase of the signal. However, only those frequency components of $\Delta a$ around the carrier frequency $f_\mathrm{c}$ contribute to the phase noise at low offset frequencies. As the self-heterodyne beating shifts the phase noise information to $f_\mathrm{c}$ where the amplitude noise of a laser is extremely low, this effectively isolates the laser RIN from entering the phase extraction process. We can therefore approximate the analytic signal by
\begin{equation}
\mathcal{H}\Delta P\approx 2iq_1^*q_2\exp(2\pi i f_0 \tau)\exp(-2\pi i f_\mathrm{c} t)(1+\Sigma a)\exp(i\Delta\phi)
\label{Eq_Hilbert_P}
\end{equation}
\end{widetext}
and the phase can be extracted as
\begin{equation}
\varphi=\Delta\phi-2\pi f_\mathrm{c} t+2\pi f_0 \tau+\mathrm{Arg}[iq_1^*q_2]
\end{equation}
After that, $\nu(\tau)$ is calculated from the time derivative of $\phi$, which can be approximated with a finite time difference:
\begin{equation}
\nu(\tau)=\frac{\dot {\varphi}}{2\pi}=-f_\mathrm{c}+\frac{\dot{ \delta\phi}(t)-\dot{\delta\phi}(t-\tau)}{2\pi}
\end{equation}
From here, the spectral density of $\nu(\tau)$ measured by a single BPD can be estimated and then used to recover $S_\nu(f)$ with the processing gain from Eq. (\ref{Eq_Gf_simple}). The results with cross-correlation can be further found in Section 3.4.

\subsection{Resolution bandwidth}
To compute $S_\nu(\tau)$ using the Wiener-Khinchin theorem, an infinite length of $\nu(\tau)$ would be required to complete the Fourier transform accurately. Using a limited amount of data leads to a finite resolution bandwidth, which will distort the measured frequency noise. Segmenting the data into shorter sections has a similar effect. Below we derive the modified PSD estimate and the corresponding $G(f)$ for the finite resolution bandwidth case.

In the calculation, $S_\nu(\tau)$ is estimated from the Fourier coefficients of the gated signal:
\begin{equation}
S_{\nu(\tau),\mathrm{gated}}(f)=\frac{|\hat{\nu}_\mathrm{gated}(\tau,f)|^2}{\int_{-\infty}^\infty w(\tau')^2\mathrm{d}\tau'}
\end{equation}
where $w(\tau')$ is the window function for gating and $\hat{\nu}_\mathrm{gated}(\tau,f)$ is the Fourier coefficient of the gated signal:
\begin{equation}
\hat{\nu}_\mathrm{gated}(\tau,f)=\int_{-\infty}^\infty \nu(\tau,t=\tau')w(\tau')\exp(2\pi i f \tau')\mathrm{d}\tau'
\end{equation}
Viewing $\nu(\tau)$ as random variables, the expectation of $S_{\nu(\tau),\mathrm{gated}}$ reads
\begin{widetext}
\begin{align}
\mathrm{E}[S_{\nu(\tau),\mathrm{gated}}(f)]&=\frac{\int_{-\infty}^{\infty}\int_{-\infty}^{\infty} \mathrm{E}[\nu(\tau,t=\tau')\nu(\tau,t=\tau'+\tau'')]w(\tau')w(\tau'+\tau'')\mathrm{e}^{2\pi if\tau''}\mathrm{d}\tau'\mathrm{d}\tau''}{\int_{-\infty}^\infty w(\tau')^2\mathrm{d}\tau'}\nonumber\\
&=\frac{\int_{-\infty}^{\infty}\mathrm{d}\tau''\left<\nu(\tau,0)\nu(\tau,\tau'')\right>\mathrm{e}^{2\pi if\tau''}\int_{-\infty}^{\infty} \mathrm{d}\tau' w(\tau')w(\tau'+\tau'')}{\int_{-\infty}^\infty w(\tau')^2\mathrm{d}\tau'}\nonumber\\
&=\int_{-\infty}^{\infty}\mathrm{d}\tau''\left<\nu(\tau,0)\nu(\tau,\tau'')\right>\mathrm{e}^{2\pi if\tau''} w_2(\tau'')\label{Eq_gated_PSD_TD}\\
&=\int_{-\infty}^{\infty}S_{\nu(\tau)}(f-f')\hat{w_2}(f')\mathrm{d}f'\label{Eq_gated_PSD_FD}
\end{align}
\end{widetext}
where we introduced $w_2(\tau'')$ as the normalized autocorrelation of $w$ and its associated Fourier transform $\hat{w_2}(f')$:
\begin{equation}
w_2(\tau'')=\frac{\int_{-\infty}^{\infty}w(\tau')w(\tau'+\tau'')\mathrm{d}\tau'}{\int_{-\infty}^\infty w(\tau')^2\mathrm{d}\tau'},\ \ \ \ w_2(0)=1
\end{equation}
\begin{equation}
\hat{w_2}(f')=\int_{-\infty}^{\infty}w_2(\tau'')\exp(2\pi if'\tau'')\mathrm{d}\tau''
\end{equation}
From Eq. (\ref{Eq_gated_PSD_FD}), it can be seen that the effect of gating the signal results in convolving $S_{\nu(\tau)}$ with $\hat{w_2}$, which is equivalent to filtering the frequency domain trace of $S_{\nu(\tau)}$ with a response function of $w_2$ in the time domain from the viewpoint of Eq. (\ref{Eq_gated_PSD_TD}).

To see how this filtering of $S_{\nu(\tau)}$ impacts signal processing, we rewrite Eq. (\ref{Eq_Snutau_simple}) as
\begin{equation}
S_{\nu(\tau)}(f)=4\sin^2(\pi f\tau)S_\nu(f)=[2-2\cos(2\pi f \tau)]S_\nu(f)
\end{equation}
If $S_\nu(f)$ varies slowly within the MZI FSR scale ($1/\tau$), the gating filter only affects the term within the brackets. For the rectangular window used here, $w_2$ becomes a triangular window, and $S_{\nu(\tau),\mathrm{gated}}$ can be found as
\begin{equation}
S_{\nu(\tau),\mathrm{gated}}(f)\approx[2-2(1-\tau \mathrm{RBW})^+\cos(2\pi f \tau)]S_\nu(f)
\label{Eq_Snutau_gated}
\end{equation}
where $z^+=\max(0,z)$ is the ramp function and RBW is the resolution bandwidth of the rectangular window (equal to the reciprocal of its temporal width). The associated processing gain becomes
\begin{equation}
G_\mathrm{gated}(f)=\frac{1}{2-2(1-\tau \mathrm{RBW})^+\cos(2\pi f \tau)}
\label{Eq_Gf_gated}
\end{equation}
We note that the divergence of $G(f)$ is no longer present in Eq. (\ref{Eq_Gf_gated}) for any finite RBW, which can be explained as a spectral leakage of noise from other offset frequencies to integer multiples of MZI FSR. If the RBW is larger than one MZI FSR such that $\tau\mathrm{RBW}>1$, then Eq. (\ref{Eq_Snutau_gated}) and Eq. (\ref{Eq_Gf_gated}) indicate that the fringe pattern is completely averaged out by the filtering. In this case the systems work in the same way as an incoherent detection setup. 

\subsection{Suppression of independent noise with cross-correlation}
While the optical signals are converted to RF signal at the BPDs, technical BPD noise (usually characterized by its noise equivalent power) will also be present in the output and is dominant in the current measurement system. This increases the phase noise of the output and limits the noise floor of the measurement without cross-correlation. We model this technical noise by adding noise terms, $\varphi_\mathrm{BPD,1}$ and $\varphi_\mathrm{BPD,2}$, for the extracted phase:
\begin{equation}
\varphi_{1}=\varphi+\varphi_\mathrm{BPD,1},\ \ \ \ \varphi_{2}=\varphi+\varphi_\mathrm{BPD,2}
\end{equation}
The noise will be transferred to the frequency signal:
\begin{equation}
\nu_1(\tau)=\nu(\tau)+\nu_\mathrm{BPD,1},\ \ \ \ \nu_2(\tau)=\nu(\tau)+\nu_\mathrm{BPD,2}
\end{equation}
where $\nu_\mathrm{BPD,1}$ and $\nu_\mathrm{BPD,2}$ are the noise terms after time-domain difference. In this case, calculating the Fourier coefficient of $\nu(\tau)$ leads to
\begin{equation}
\hat{\nu}_1(\tau,f)=\int_{-\infty}^\infty [\nu(\tau,t=\tau')+\nu_\mathrm{BPD,1}]w(\tau')\mathrm{e}^{2\pi i f \tau'}\mathrm{d}\tau'
\end{equation}
Assuming $\nu(\tau)$ and $\nu_\mathrm{BPD,1}$ are independent, the calculated PSD for $\nu_1(\tau)$ becomes
\begin{equation}
S_{\nu(\tau),1}(f)=S_{\nu(\tau),\mathrm{gated}}(f)+S_\mathrm{BPD,1}
\end{equation}
where $S_\mathrm{BPD,1}$ is the gated PSD for $\nu_\mathrm{BPD,1}$ noise and determines the measurement floor using only a single BPD.

To remove the BPD technical noise, both BPD outputs are used and cross-correlated to suppress the contribution of $\nu_\mathrm{BPD,1}$ and $\nu_\mathrm{BPD,2}$. The correlated estimate of $S_{\nu(\tau)}(f)$ is the product of two Fourier coefficients originating from different BPDs:
\begin{equation}
S_{\nu(\tau),\mathrm{corr}}(f)=\frac{\hat{\nu}_1(\tau,f)\hat{\nu}_2^*(\tau,f)}{\int_{-\infty}^\infty w(\tau')^2\mathrm{d}\tau'}
\end{equation}
Assuming $\nu(\tau)$, $\nu_\mathrm{BPD,1}$ and $\nu_\mathrm{BPD,2}$ are all independent, it can be readily shown that
\begin{equation}
\mathrm{E}[S_{\nu(\tau),\mathrm{corr}}(f)]=S_{\nu(\tau),\mathrm{gated}}(f)
\end{equation}
and includes only the contributions from laser noise. However, BPD noise adds randomness to the correlation and increases the variance of $S_{\nu(\tau),\mathrm{corr}}(f)$. This effect will be more obvious at high offset frequencies when technical phase noise from the BPD is converted to larger frequency noise (see Section 4.3), or at low offset frequencies while using a short delay line (where $G(f) \gg 1$). This can be improved by averaging over $N$ segments of data, which lowers the standard error of the mean by $\sqrt{N}$ times and therefore improves the signal-to-noise ratio by $\sqrt{N}$.

% Characterization
\section{Characterization of the setup}

\subsection{RIN suppression}
\begin{figure*}
\includegraphics[width=132mm]{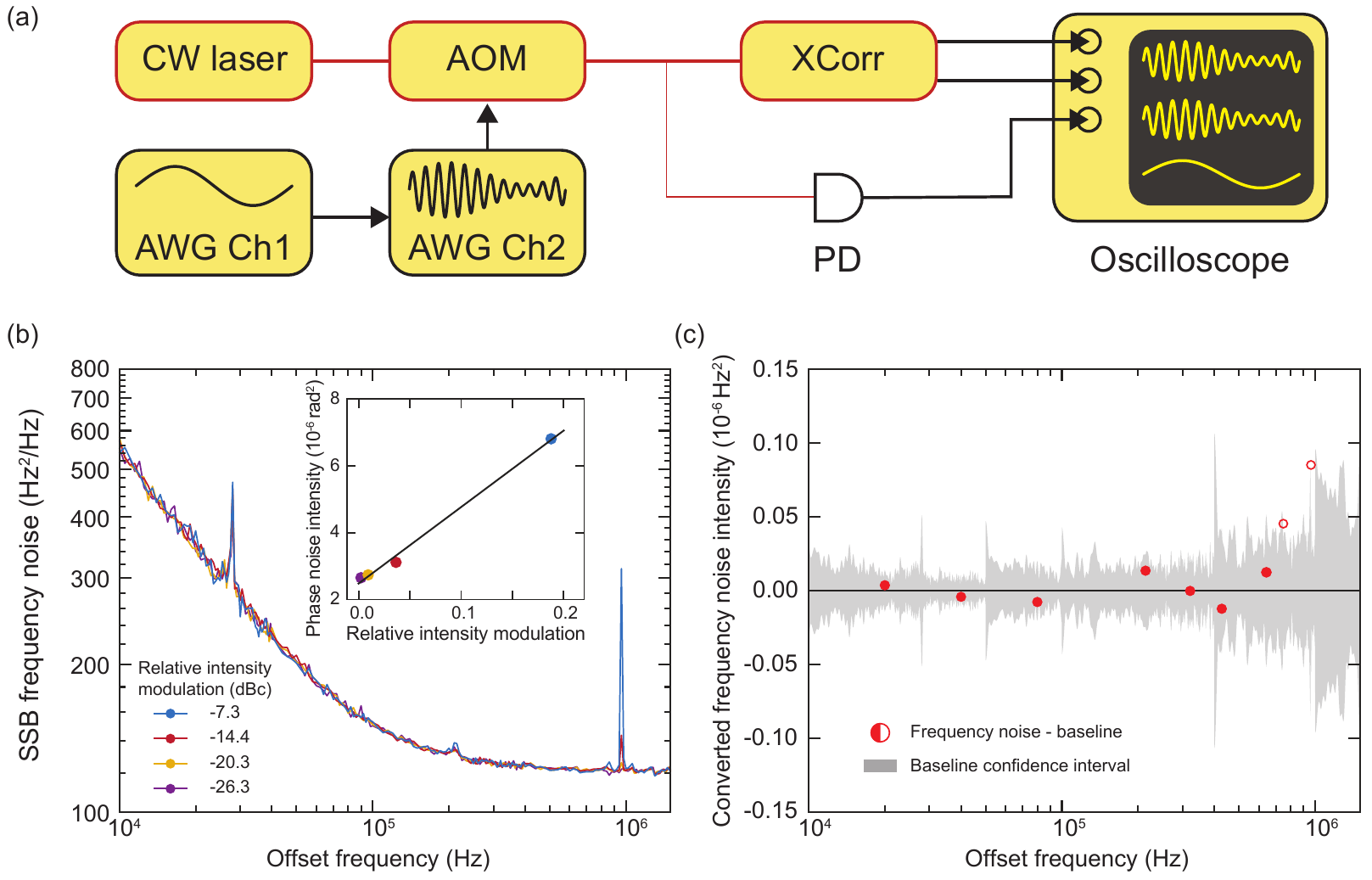}
\caption{(a) Experimental setup for RIN suppression characterization. CW laser: continuous wave laser, AOM: acousto-optical modulator, XCorr: cross-correlator [as in Fig. \ref{FIG1}(a)], AWG: arbitrary waveform generator, PD: photodetector, OSC: oscilloscope. (b) SSB laser frequency noise spectra under different modulation intensity at 963 kHz offset frequency. Inset: Phase noise intensity at 963 kHz as a function of modulation intensity. The solid black line is the linear fitting. (c) Frequency dependence of RIN conversion ratio. The shaded area stands for the confidence interval (99\% confidence probability) of the frequency noise intensity without power modulation. The frequency noise intensity baseline has been subtracted for comparison. The red circles are plotted as the converted frequency noise intensity (measured frequency noise intensity minus baseline) at different offset frequency corresponding to the power modulation frequency. The solid (hollow) circles indicate that the converted frequency noise intensity is inside (outside) the shaded area.
}
\label{FIG2}
\end{figure*}

As noted in Section 3.2, the frequency-shifting process of the self-heterodyne setup effectively isolates RIN from coupling into the measured frequency noise. In order to characterize the RIN suppression performance of the cross-correlator, we measured the RIN conversion ratio with a setup shown in Fig. \ref{FIG2}(a). The measurement setup consists of an ECL (RIO ORION 1550 nm laser module) modulated by an AOM to generate an artificial RIN signal. The AOM carrier generated from the arbitrary waveform generator is amplitude-modulated by a single-tone sine wave with manually configured frequency and modulation depth. The carrier frequency is selected as the optimal modulation frequency of this AOM (here 55 MHz) to minimize amplitude-phase coupling during the modulation process. By measuring the frequency noise with and without the power modulation using the cross-correlator, the RIN conversion ratio can be determined. The modulation intensity is calculated from the modulation depth and calibrated by tapping 10\% of the modulated laser before the cross-correlator. All signals are recorded by the aforementioned high-speed oscilloscope.  

Empirically, the RIN conversion to measured frequency noise can be described by
\begin{equation}
\tilde{S}_{\nu(\tau)}(f)=S_{\nu(\tau),\mathrm{corr}}(f)+f^2\alpha\times \mathrm{RIN}(f)
\label{Eq_RIN_emp_model}
\end{equation}
where $\tilde{S}_{\nu(\tau)}(f)$ is the measured frequency noise just before the processing gain, including RIN contributions, $\alpha$ is a proportionally constant that converts RIN to phase noise, and the extra $f^2$ factor further converts phase noise to frequency noise. For the reconstructed laser noise the above equation becomes
\begin{equation}
\tilde{S}_\nu(f)=S_\nu(f)+G(f)f^2\alpha\times \mathrm{RIN}(f)
\label{RIN}
\end{equation}
The single-tone modulations used in the actual measurements could not be quantified by a spectral density. Instead, the frequency noise intensity can be recovered from
\begin{equation}
P_{\nu}(f_\mathrm{AM}) = S_{\nu}(f_\mathrm{AM})\times\text{RBW}(f_\mathrm{AM})
\end{equation}
where $f_\mathrm{AM}$ is the modulation frequency. Similarly, phase noise intensity is related to the frequency noise intensity by
\begin{equation}
P_{\phi}(f_\mathrm{AM}) = \frac{1}{f_\mathrm{AM}^2}P_{\nu}(f_\mathrm{AM})
\end{equation}
By comparing $P_{\phi}(f_\mathrm{AM})$ against the amplitude modulation intensity, $\alpha$ can be extracted through a linear fitting process.

The measured $\tilde{S}_\nu(f)$ under different modulation intensity at 963 kHz offset frequency are shown in Fig. \ref{FIG2}(b). The noise spurs at 29 kHz come from the ECL itself, which also appear in Fig. \ref{FIG5} and Fig. \ref{FIG6}. The inset of Fig. \ref{FIG2}(b) shows the linear fitting between the phase noise intensity and the modulation intensity. Here, the RIN to measured laser phase noise conversion ratio [$G(f)\alpha$] at 963 kHz (i.e., the slope of the fitting) is -46.4 $\text{dBrad}^2$ (2.29 $\times10^{-5}$ rad$^2$). From here, $\alpha$ can be found as -40.4 $\text{dBrad}^2$ (9.16 $\times10^{-5}$ rad$^2$).

As shown in Fig. \ref{FIG2}(c), a -20 dBc modulation intensity is selected for all measurements at different modulation frequencies, which is sufficient as an overestimation for RIN of a normal laser. The intrinsic laser frequency noise may obscure the presence of weakly-coupled RIN, and becomes a noise floor for the RIN conversion measurement. The frequency noise intensity without laser power modulation (i.e., the baseline) is measured 10 times and a 99\% possibility confidence interval (shaded area) for the RIN conversion signal is given by assuming that the measured frequency noise intensity is normally distributed. When amplitude modulation is applied, the frequency noise intensity (red circles) at the corresponding offset frequency is calculated and compared with the baseline. A measured intensity outside the confidence interval indicates significant conversion of RIN. RIN conversion has been tested at 20 kHz, 40 kHz, 80 kHz, 214 kHz (equal to the MZI FSR $1/\tau_0$, where $\tau_0$ = 4.67 ms is the delay time of the 1-km-long fiber), 321 kHz, 428 kHz, 642 kHz, 749 kHz, and 963 kHz. The modulation frequencies over 100 kHz match the integer and half-integer multiples of MZI FSR, corresponding to local maximums and minimums of $G(f)$, respectively. The measured frequency noises with artificial RIN coupling only fall outside of the confidence interval at 749 kHz and 963 kHz, which suggests that the RIN suppression is high enough for lasers at offset frequency lower than 1 MHz. We note that the $G(f)$ will amplify the RIN conversion at offset frequencies equal to integer multiples of the MZI FSR. However, as shown by the experimental data, the overall RIN conversion is not important in most cases.

Below we present a model for the origin of RIN coupling by considering signal leakage at the AOM within the cross-correlator [Fig. \ref{FIG1}(a)]. We assume that a small portion of 0$^\text{th}$ order light is leaked into the 1$^\text{st}$ order port at the AOM, which is supported by experimental observations. The net effect is that the amplitude for the frequency-shifted arm should be modified as
\begin{widetext}
\begin{align}
\tilde{A_1}(t) &=\frac{1}{\sqrt{2}}\exp(-2\pi if_0 t)\exp(2\pi i f_\mathrm{c} t)[1+\delta a(t)]\exp[-i\delta \phi(t)]\nonumber\\
&+\frac{\epsilon}{\sqrt{2}}\exp(-2\pi if_0 t)[1+\delta a(t)]\exp[-i\delta \phi(t)]
\end{align}
where $\epsilon$ is a complex number that represents the leakage amplitude. The amplitude on the other arm $A_2$ remains the same. The BPD power now reads, keeping only the signals oscillating near frequency $f_\mathrm{c}$,
\begin{align}
\Delta \tilde{P} &= 2\mathrm{Re}[iq_1^*q_2\exp(2\pi i f_0 \tau)\exp(-2\pi i f_\mathrm{c} t)(1+\Sigma a)\exp(i\Delta \phi)] \nonumber\\
&+\mathrm{Re}[\Delta|q|^2 \epsilon \exp(-2\pi i f_\mathrm{c} t)(1+2\delta a(t))]
\end{align}
where the exponential of phase noise is linearized for convenience. The extra term results from the same-arm beating detected at the BPD. The analytic signal is given by, up to first order of $\epsilon$ and $\Delta\phi$,
\begin{equation}
\mathcal{H}\Delta P\approx 2iq_1^*q_2\exp(2\pi i f_0 \tau)\exp(-2\pi i f_\mathrm{c} t)\left[1+\Sigma a+i\Delta\phi+\frac{\Delta|q|^2|\epsilon|}{|q_1q_2|}\exp(i\theta_\mathrm{RIN})\left(\frac{1}{2}+\delta a(t)\right)\right]
\end{equation}
\end{widetext}
Here $\theta_\mathrm{RIN}$ is a phase angle that couples amplitude to phase:
\begin{equation}
\theta_\mathrm{RIN}=2\pi i f_0 \tau +\mathrm{Arg}[\epsilon]-\mathrm{Arg}[iq_1^*q_2]
\end{equation}
Performing phase extraction and time difference leads to
\begin{equation}
\tilde{\nu}(\tau)=\nu(\tau)+\frac{\Delta|q|^2|\epsilon|}{|q_1q_2|}\sin(\theta_\mathrm{RIN})\frac{\mathrm{d}\delta a}{2\pi \mathrm{d}t}
\end{equation}
For the worst case (maximal coupling) $\sin(\theta_\mathrm{RIN})=\pm 1$, and in the case when laser amplitude and frequency noise are independent, calculating the PSD gives
\begin{equation}
\tilde{S}_{\nu(\tau)}(f)=S_{\nu(\tau),\mathrm{corr}}(f)+f^2\frac{(\Delta|q|^2)^2|\epsilon|^2}{|q_1|^2|q_2|^2}\frac{\mathrm{RIN}(f)}{4}
\end{equation}
Comparing with the empirical model Eq. (\ref{Eq_RIN_emp_model}) gives
\begin{equation}
\alpha = \frac{(\Delta|q|^2)^2}{4|q_1|^2|q_2|^2}|\epsilon|^2
\end{equation}
and is directly proportional to the leaked power at the AOM.

\begin{figure}
\includegraphics[width=\linewidth]{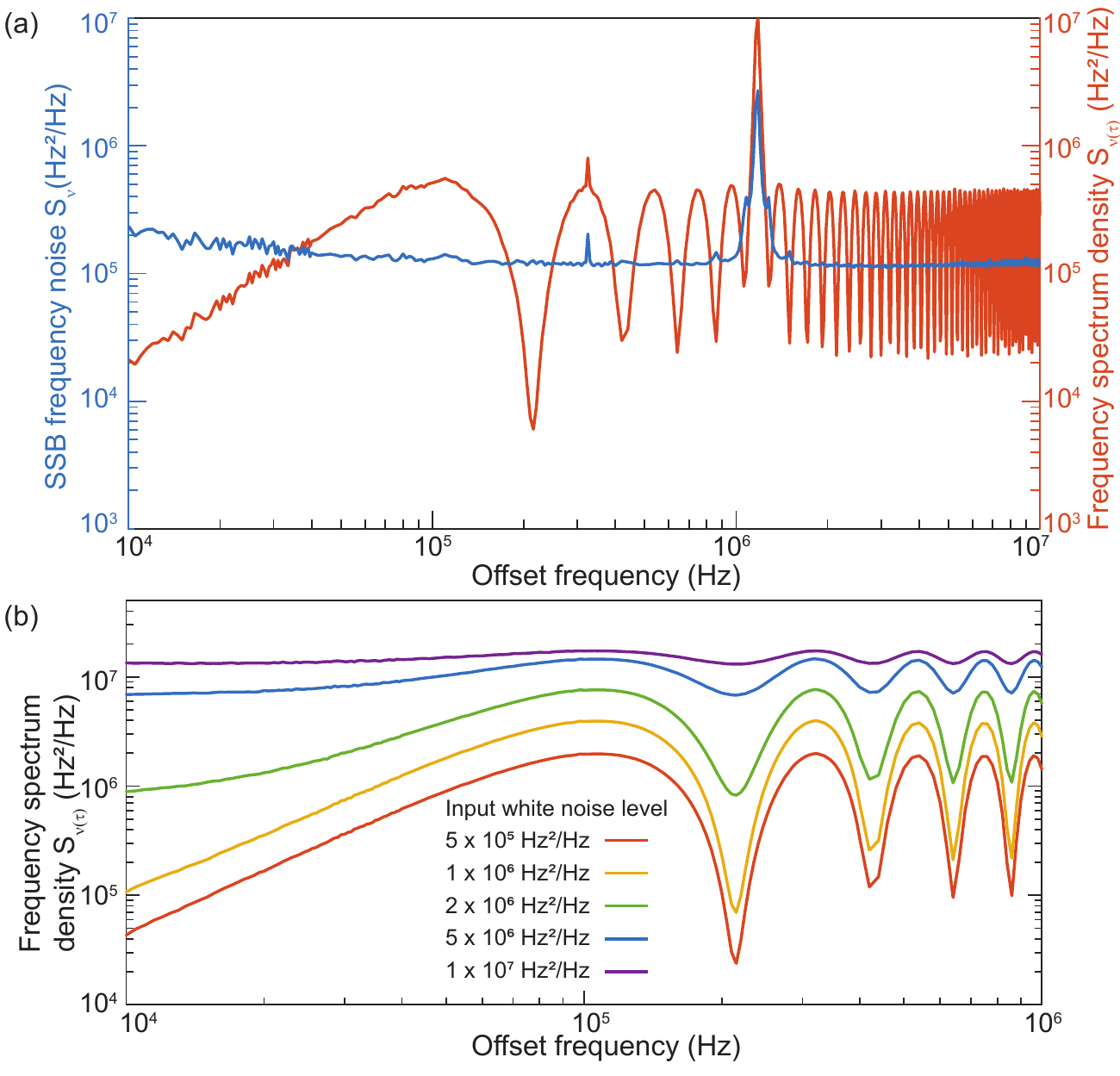}
\caption{(a) SSB frequency noise and frequency spectral density of a DFB laser. 
(b) Simulated $S_{\nu(\tau)}(f)$ output of the system when a laser with high white frequency noise is used as input. The white noise levels are, from bottom to top: $5 \times 10^5$ Hz$^2$/Hz, $1 \times 10^6$ Hz$^2$/Hz, $2 \times 10^6$ Hz$^2$/Hz, $5 \times 10^6$ Hz$^2$/Hz, $1 \times 10^7$ Hz$^2$/Hz.}
\label{FIG3}
\end{figure}

\subsection{Dynamic range}%DFB laser noise

\begin{figure}
\includegraphics[width=\linewidth]{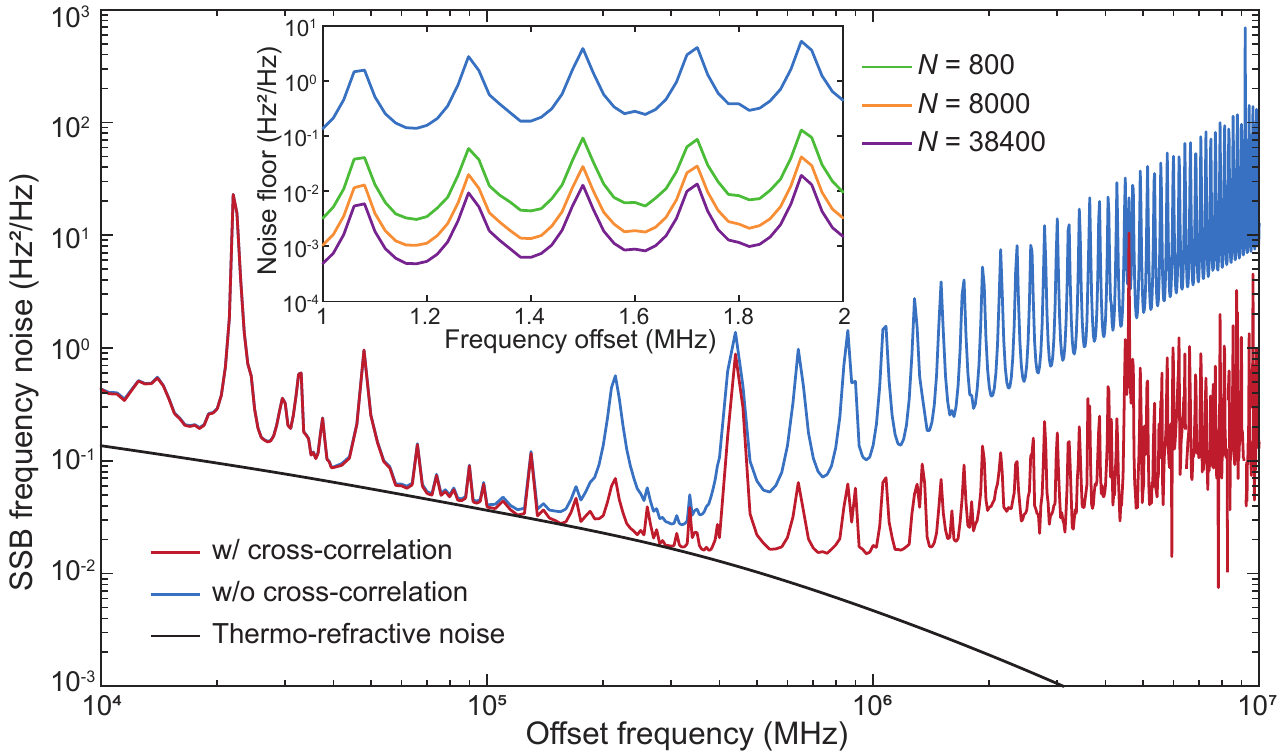}
\caption{SSB frequency noise of the DFB laser with SIL measured by the self-heterodyne method with and without cross-correlation. The deviation between them is due to the BPD noise as illustrated in Fig. \ref{FIG1}. The cross-correlator extracts the common mode SIL laser noise and suppresses the individual BPD noise to reach a 0.01 $\text{Hz}^2$/Hz noise floor at 1 MHz offset frequency. The simulated thermo-refractive noise is plotted in black for comparison. 
Inset: the measured frequency noise error bar is interpreted as the measurement system noise floor here. With increasing averaging segment number $N$, the noise floor is reduced by $\sqrt{N}$. The blue trace is the same as the main figure.
}
\label{FIG4}
\end{figure}

To quantify the upper limit of noise the COSH setup can measure, we use a DFB laser for characterization. Limited by the laser cavity length and reflectivity of the output facet, the noise of a DFB laser can reach the level of 1 MHz, corresponding to a coherence length shorter than the 1-km-long fiber delay line. However, the measurable noise is not directly related by the delay line, but only limited by the carrier frequency (see below).

Measurement results for a free-running DFB laser are shown in Fig. \ref{FIG3}(a). The SSB frequency noise of the DFB laser is around $1.2 \times 10^5$ Hz$^2$/Hz at high offset frequencies, corresponding to a 0.75 MHz Lorentzian linewidth and a 267 m coherence length in fiber (assuming a fiber refractive index of 1.5). As the BPD noise is far less than the DFB laser noise, performing cross-correlation on the data provides negligible improvement. Note that the noise peak at 1.2 MHz is from the DFB laser. Combined with the estimate for the noise floor (see Section 4.3), the setup could reach over 70 dB dynamic range for frequency noise measurements.

To understand how the system behaves for higher laser frequency noise, simulations have been performed and the results are collected in Fig. \ref{FIG3}(b). Increasing the laser frequency noise above $1 \times 10^6$ Hz$^2$/Hz leads to a visible decrease of fringe contrast before multiplying $G(f)$. This can be attributed to the wide broadening of the carrier signal $\exp(2\pi i f_\mathrm{c} t)$. The components that are separated more than $f_\mathrm{c}$ from the carrier cross into the negative-frequency domain and will be reflected by the Hilbert transform. If such contributions are significant, Eq. (\ref{Eq_Hilbert_P}) is invalidated, and the laser frequency noise can no longer be reliably recovered. Choosing an AOM with higher modulation frequency and increasing the sampling rate for the oscilloscope could increase the upper noise limit at the expense of measurement time or memory. We note that fringes are still visible even if the noise level exceeds the MZI FSR (i.e. the laser coherent length is shorter than the delay length), unless the resolution bandwidth is chosen to exceed the MZI FSR [as given by $G(f)$].

\subsection{Noise Floor}%DFB laser with self-injection-locking (SIL)

Finally, the noise floor of the COSH method is verified using a DFB injection locked to a high-Q resonator. The laser linewidth coming from the compound laser-resonator system can be greatly suppressed \cite{dahmani1987frequency, hjelme1991semiconductor, kondratiev2017self} and have demonstrated record linewidth levels in integrated photonics platforms comparable to fiber lasers \cite{li2021reaching}. 

Here, the aforementioned DFB laser has been self-injection-locked to a 7-m long, ultra-high-Q on-chip resonator (with an intrinsic Q factor of 150 million). The SSB frequency noise of this lab-built laser is then measured by the setup and results are shown in Fig. \ref{FIG4}.
Compared with a 1.4-m long spiral resonator \cite{li2021reaching}, larger mode volume further suppresses the thermo-refractive noise (TRN) and reaches 0.041 Hz$^2$/Hz at 100 kHz offset frequency. The numerically simulated TRN is also plotted for comparison. 
The lowest measured frequency noise is $0.015 \pm 0.002$ Hz$^2$/Hz at 1 MHz offset frequency, comparable with the previous work \cite{li2021reaching}. 

The power spectrum density given by a single BPD output frequency without cross-correlation is also illustrated in Fig. \ref{FIG1}d. Since BPD technical noise is approximately white when characterized as phase noise, the independent BPD noise contribution to measured frequency noise scales as $f^2$ and is more apparent at high offset frequencies, which is confirmed by comparing the two traces in Fig. \ref{FIG4}. The spurs in the single-BPD trace are the BPD technical noise amplified by $G(f)$ and are also an indication that the BPD noise has significant contributions. 
By using cross-correlation and averaging over $N$ segments of data, the noise contribution can be reduced by $\sqrt{N}$. Here, for the 20 kHz resolution bandwidth used for high offset frequencies, we have $N=38400$, and the signal-to-noise ratio is improved by 22.9 dB. As shown in the inset of Fig. \ref{FIG4}, the error bar of measured frequency noise (standard deviation of frequency noise from multiple segments) is interpreted as the measurement setup noise floor here, and larger N (proportional to the overall data length used) leads to a lower noise floor.
Assuming 0.05 mW optical input power at the BPD, the technical noise is equivalent to 0.10 Hz$^2$/Hz at 1 MHz offset frequency, consistent with the blue trace in Fig. \ref{FIG4} inset. With $N=38400$, the noise floor with cross-correlation is suppressed to be 0.0005 Hz$^2$/Hz, consistent with the purple trace in the inset. At MZI FSR frequencies, the noise floor is enhanced but remains below 0.01 Hz$^2$/Hz around 1 MHz offset frequency.
The spurs of measured SSB frequency noise at higher than 1 MHz offset frequency is significantly higher than the noise floor and are believed to originate from the residual RIN amplified by $G(f)$.

\section{Discussion}

\subsection{Environmental noise coupling}
\begin{figure}[t]
\includegraphics[width= \linewidth]{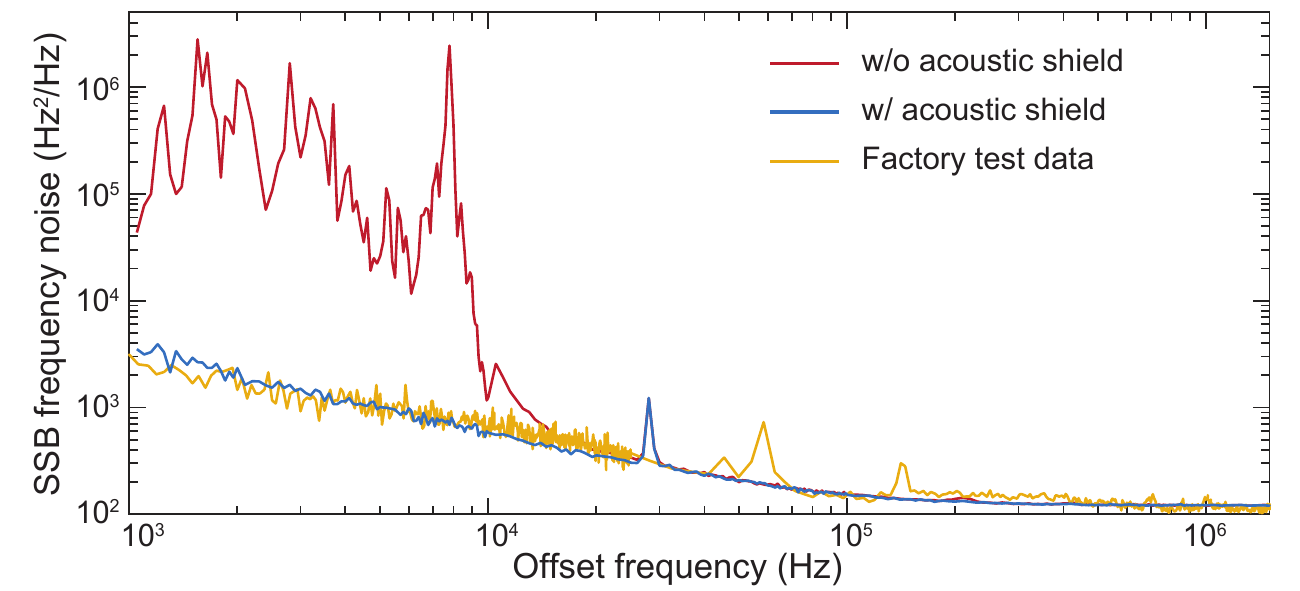}
\centering
\caption{Typical SSB frequency noise of a commercial ECL obtained using cross-correlator with/without an acoustic shield. The yellow trace is the test data as provided by the manufacturer.}
\label{FIG5}
\end{figure}
Acoustic noise may be present in the external environment and can couple to the measured SSB frequency noise through the fiber delay of the modified MZI. To minimize environmental perturbations, the optical section of the measurement setup is acoustically shielded with a foam box. To calibrate the external noise isolation from the shield box, we have applied additional acoustic noise in the environment and measured the ECL frequency noise. As is shown in Fig. \ref{FIG5}, the measured SSB frequency noise is smooth and no peaks can be found below 10 kHz offset frequency, compared to the case without the acoustic shield. Above 10 kHz, the effect of environment noise is not evident for the noise measurement system.
Meanwhile, the frequency noise measured with acoustic shield is consistent with the laser manufacturers specification sheet.

\subsection{Fiber delay length}
A major drawback of the current setup is the decrease of sensitivity at integer multiples of MZI FSR, where the frequency noise destructively interferes and $G(f)$ reaches its maximum. These frequencies can be adjusted by changing the fiber delay length. If the destructive interference is undesired over a wide offset frequency range, the fiber delay length should be short enough such that the first MZI FSR appears outside the frequency range. For example, a 10 meter fiber (with MZI FSR equal to approximately 20 MHz) ensures that no fringes appear below 20 MHz offset frequency.

\begin{figure}[t]
\includegraphics[width=\linewidth]{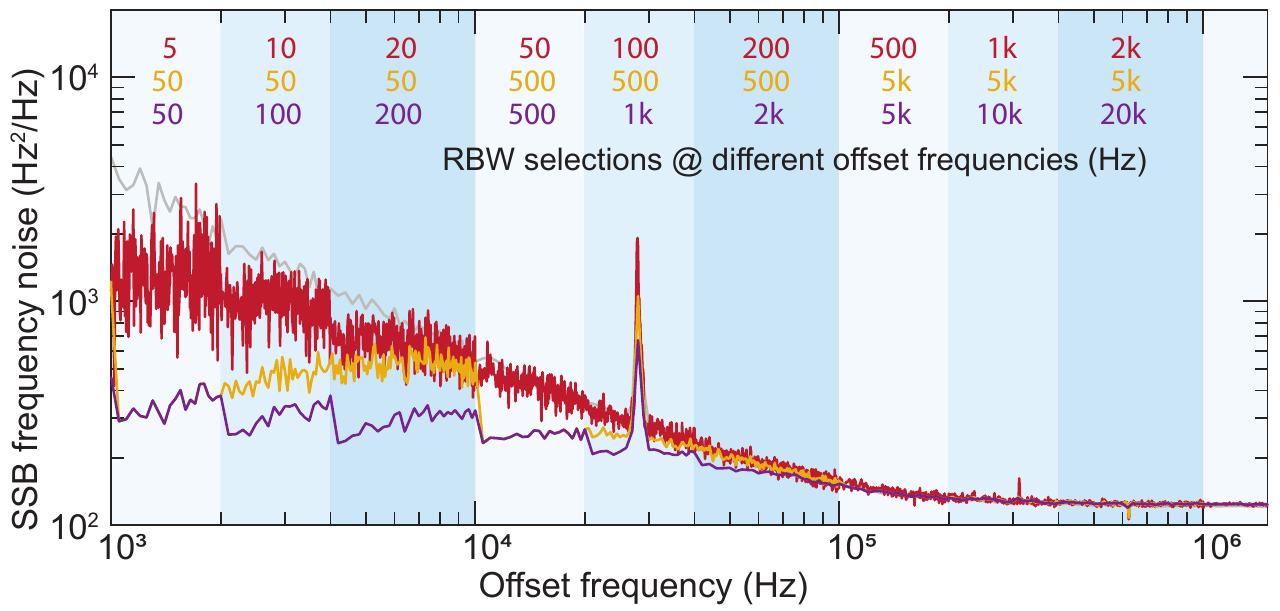}
\caption{SSB frequency noise of the ECL calculated with different RBW configurations using 15-meter-long fiber delay line are plotted as red, yellow and purple traces.
The RBWs chosen at different offset frequency intervals (marked by shading) are shown at the top with colors in accordance with traces.
The gray line using 1-km-long fiber delay line is plotted as a reference and its RBW configurations are the same as the purple trace.}
\label{FIG6}
\end{figure}

A major disadvantage of using a short fiber delay length is the large systematic error of low-frequency noise. To demonstrate this, the 1 km delay line in the MZI is substituted with a 15 meter fiber and the ECL noise is measured. The results are presented in Fig. \ref{FIG6}. We note that there are frequency noise discontinuities when the RBW changes. For the purple trace in Fig. \ref{FIG6}, the calculated frequency noise ``jumps'' at 20 kHz, 40 kHz and 100 kHz, where different RBWs are chosen for the offset frequency intervals on both sides. On the other hand, the data measured using 1 km delay length (grey trace) gives a continuous result and is consistent with the noise data from the laser manufacturers specification sheet.

The strong dependence of calculated noise on the RBW can be attributed to the non-white frequency noise spectrum of the laser. The gating filter acts differently on the spectrum compared to the white case and invalidates Eq. (\ref{Eq_Snutau_gated}). By using smaller RBWs, the calculated noise becomes closer to the true laser noise, as seen from the red trace in Fig. \ref{FIG6}.

\section{Conclusion}

In this paper we have demonstrated a correlated self-heterodyne (COSH) method to measure laser frequency noise with 0.01 $\text{Hz}^2$/Hz noise floor and high RIN rejection quantified by the coupling coefficient $\alpha = -40.4$ dBrad$^2$. Commercial ECL and DFB lasers with/without self-injection-locking are used to verify the performance. The cross-correlation noise floor is limited by the coupled RIN amplified by the processing gain $G(f)$ as well as residual BPD technical noise.

The setup described here can be further reconfigured to meet specific measurement requirements. For example, the RIN suppression can be further enhanced by using an AOM with higher ${0}^\text{th}$ order to ${1}^\text{st}$ order isolation. Temperature controllers can be installed to the 50/50 couplers to adjust the coupling ratio precisely and balance the MZI arms. On the other hand, if the expected laser noise frequency is high, then cross-correlation is not necessary and the memory depth of the oscilloscope can be decreased accordingly. A conventional PD can be used in place of a BPD if RIN is not a concern. Overall, the specific measurement setup and parameters introduced in Section 2.1 are targeted towards ultra-low-noise laser measurement mainly at high offset frequencies, while the basic principle remains universal.

The authors acknowledge Q. Yang and Y. Lai for helpful discussions. This project was supported by the Defense Advanced Research Projects Agency (DARPA) under the LUMOS program (HR001-20-2-0044); and by the Air Force Office of Scientific Research (FA9550-18-1-0353).

Data underlying the results presented in this paper may be obtained from the authors upon reasonable request. Code that processes oscilloscope data has been released as a Python package (\verb|pycosh|) and can be found at the Python Package Index (https://pypi.org/).

\end{document}